# $3_{10}$ and π-helices: Stochastic Events on Sequence Space; Reasons and Implications of their Accidental Occurrences across Protein Universe.

Param Priya Singh, Anirban Banerji[*]
Bioinformatics Centre, University of Pune, Pune-411007, Maharashtra, India.
Contact details: anirbanab@gmail.com

## Abstract

Occurrences of $3_{10}$ and π-helices in proteins are rare, but are hard to ignore. Despite numerous studies regarding compositional and energetic profile of $3_{10}$ and π-helices in particular cases, objective and general insights about reasons and patterns of occurrence of these (rather unstable) structures in proteins are still unclear. To understand the reasons behind the existence of $3_{10}$ and π-helices, one needs to unambiguously describe the general nature of their occurrence profile on primary structures, at the first place. Considering all available non-redundant protein structures across different structural classes, present study identified the probabilistic characteristics that describe several facets of the occurrence of $3_{10}$ and π-helices in proteins. Occurrence profile of $3_{10}$ and π-helices revealed that, their presence follows Poisson flow on the primary structure; implying that, their occurrence profile is rare, random and accidental. Structural class-specific statistical analyses of sequence intervals between consecutive occurrences of $3_{10}$ and π-helices revealed that these could be best described by gamma and exponential distributions, across structural classes. Comparative study of normalized percentage of non-glycine and non-proline residues in $3_{10}$, π and α-helices revealed a considerably higher proportion of $3_{10}$ and π-helix residues in disallowed, generous and allowed regions of Ramachandran map. Probe into these findings in the light of evolution suggested clearly that $3_{10}$ and π-helices should appropriately be viewed as evolutionary intermediates on long time scale, for not only the α-helical conformation but also for the 'turns', equiprobably. Hence, accidental and random nature of occurrences of $3_{10}$ and π-helices, and their evolutionary non-conservation, could be described and explained from an invariant quantitative framework. Extent of correctness of two previously proposed hypotheses on $3_{10}$ and π-helices, have been investigated too. Alongside these, a new algorithm to differentiate between related sequences is proposed, which reliably studies evolutionary distance with respect to protein secondary structures.

## Introduction

Three types of helical secondary structures are found in proteins: α, $3_{10}$ and π-helices. They are characterized by different hydrogen bonding patterns, rise per residues and atoms contained in ring joined by hydrogen bonds. The α-helix ($3.6_{13}$ helix, with the transition of 1.5 Å along the helix axis) is characterized by $i→(i+4)$ hydrogen bonding pattern. In contrast, $3_{10}$ helices (with a transition of 2 Å along the helix axis) and π-helices ($4.4_{16}$ helix, with a transition of 2Å along the helix axis) are characterized by $i→(i+3)$ and $i→(i+5)$ hydrogen bonding patterns respectively [1]. Although α-helices are the most commonly found helical secondary structures in proteins, events of presence of the other two helices are hard to ignore. Present study explores the statistical and evolutionary patterns in the occurrence profile of $3_{10}$ and π-helices. These helices will be referred to as 'rare helices' (RH) throughout this article.

Although possibility of existence of $3_{10}$ helices in proteins was discussed eight years before Pauling proposed the structure of α-helix [2, 3], studies on different structural and evolutionary aspects of $3_{10}$ and π-helices are miniscule as compared to the same on α-helices. It is known however that $3_{10}$ and π-helices are irregular in shape, they are relatively unstable and their presence in protein backbone introduces local steric constrains. However, despite their inherent unstable conformation, case-specific (but not general) functional significances of RH have been reported in numerous cases; e.g. in motif formation [4], in protein-protein interaction [5], active site geometry [6-7], RNA binding [8], receptor binding [9] etc. Still, the reasons and implications of their occurrences in protein in general (and not in some particular cases) – is not known, to the best of our knowledge.

A systematic survey of available literature on RH reveals a lack of consensus on most of their properties. For example, to describe $3_{10}$-helices in the torsion angle space, a range of mean torsion angle (φ-ψ) values have been reported in the literature, with significant amount of variation [5-14]. On energetic perspective, while a number of theoretical and computational studies state that α-helices are energetically more stable than $3_{10}$-helices [15-19], analysis of electron spin resonance spectral data pointed at the coexistence of $3_{10}$ and α-helices [20-21]. One, however, finds it a bit irreconcilable to associate such coexistence with reports of easy melting of $3_{10}$-helices at low temperature [22]. Moving on to other aspects, although it is intuitive to expect that initiation of helix formation will be easier in case of $3_{10}$-helix than for α-helix (one less unit to consider while forming the first hydrogen bond), it has been reported [23] that it is not $3_{10}$-helices but only α-helices, β-sheets or short covalently bridged cycles (as in conotoxins or in metallothioneins), which can serve as nucleation sites for protein folding. But there again, the last assertion appears to be in sharp contrast to the results obtained from studying helix-coil transition phenomena with various peptides and proteins, where some works claimed the formation of $3_{10}$-helices as the necessary step *en-route* to α-helix formation [20, 24-25]. Investigating helix-coil transition and helix formation phenomenon on Ala-rich peptides with double-label electron spin resonance spectral data and NMR, a group of reports claimed a coexistence of $3_{10}$ and α-helices [26-27]. Claims of formation of $3_{10}$-helices as necessary precursors for α-helix formation [20, 24-25], provide an indirect plausibility to such claim. However, one also notices a stark counter-claim, asserting that the formation of $3_{10}$-helices is not a necessary prerequisite for the formation of α-helices [28]. But the root cause behind such diverse array of inconclusive findings has never been identified.

Owing to unfavorable energy due to cavity formation in the helix interior and entropic costs [7, 15, 29], π-helices are considered as unstable structures. π-helices are far less frequent than $3_{10}$-helices [30]; number of studies on them is proportionately less too. Nevertheless, one notices significant inconsistencies about various reported features of π-helices too. For example, just like the case of $3_{10}$-helices, one finds a range of reported torsion angle (φ-ψ) values with significant amount of variation for π-helices as well [15, 29, 31-33]. Although π-helices have been proved by many studies as energetically unfavorable [7, 15, 29], a circular dichroism study [34] on regular small peptides suggested that α-helical and π-helical conformations might share indistinguishable spectra. Another study, acknowledging the rare occurrence of π-helices and pivotal dependence of their stability on residue-arrangement, reported an interesting case where peptides demonstrated higher propensity to form π-helices over α-helices [35]. Finally, the assertion from a previous study [32] that conformational properties of π-helices, as obtained through in simulations of peptides or proteins, is force field-dependent - leads one to be unconvinced about many works that ostensibly report various stereochemical properties of π-helices, merely on the basis of computational investigations.

In this context it assumes significance to mention that many of the $3_{10}$ and π-helix studies involve synthetic peptides, especially α-aminoisobutyric acid (AIB). It is interesting to note that AIB is not a 'proteinogenic' amino acid; furthermore, although it is rare to be found in nature, in oligomeric form it readily forms $3_{10}$-helices [36]. Therefore, to what extent a hypothesis regarding RH stability constructed out of AIB studies is relevant in the complex domain of proteins, becomes a rather philosophical question.

Objective of the present work was twofold. On one hand, it attempted to investigate the general reasons behind existence of $3_{10}$ and π-helices across protein universe. On a related note, certain corollaries of this work helped us to ascertain the probable reasons behind the inconclusive (and at times contradictory) nature of reported results regarding $3_{10}$ and π-helix properties. On the other hand, current work attempted to test two extremely important hypotheses regarding structural and evolutionary aspects of $3_{10}$ and π-helices, as put forward by previous works. Consequently, the hypothesis that $3_{10}$-helices are possibly the intermediates in α-helix formation [20, 30, 37], and the claim that π-helices in proteins cannot be considered as a true linear group of repeating φ-ψ pairs – were tested here [38].

The study begins by analyzing the occurrence profile of RH on sequences of non-redundant protein structures. This analysis was general; hence, no special attention was paid to particular cases where RHs are reported to have certain (possible) structural and/or functional role. In the next stage, a probabilistic model was constructed to investigate the general patterns in occurrence profile of $3_{10}$ and π-helices on primary structures. Results obtained from probabilistic analysis found support from results of an exhaustive statistical survey of inter-arrival distribution profile of $3_{10}$ and π-helices, conducted on primary structures of all non-redundant proteins distributed across all the structural classes. Finally, on the third stage, effect of the presence of $3_{10}$ and π-helices on the evolution of protein secondary structures was investigated. Evolutionary examination helped us in probing the extent, to which $3_{10}$ and π-helices can be considered as possible intermediates in α-helix formation, as hypothesized in some previous works.

Since $3_{10}$-helices in helical peptides are reported to have sequence dependence [39-40], the general structural and/or functional reason(s) behind existence of $3_{10}$ and π-helices across protein universe, if any, was expected to be represented by certain tangible or latent statistical pattern(s) on the primary structures. Hence the primary goal of the present study was to describe and decipher the reasons behind hidden bias or patterns (if any), in the rare yet noticeable occurrence profile of $3_{10}$ and π-helices on primary structures; before interpreting them from both structural and evolutionary standpoint. To make certain that this goal is indeed achieved; global nature of dataset was ensured throughout the scope of the work. Obtained results suggested that $3_{10}$ and π-helices follow Poisson family of distributions in their occurrence profile on primary structures. In general terms, their occurrences could be interpreted as stochastically distributed, across all different structural classes of proteins. Based on these results, adhering strictly to the implications of Poisson distribution and structural findings, we hypothesized that, $3_{10}$ and π-helices are accidentally occurring structures, with little importance to proteins in general. Although such an assertion seemed in agreement with broad statistical trend, "nothing makes sense in biology except in the light of evolution" [41]. Hence, to put our hypothesis to stringent test, it was subjected to evolutionary findings. Results from the evolutionary study, however, neither definitely accepted nor completely rejected our hypothesis. Rather, it suggested that $3_{10}$ and π-helices should appropriately be viewed as stochastically distributed imperfect structures that are long-term evolutionary intermediates for the formation of α-helices and, equally probably, the loops.

## Results:
### $3_{10}$ and π-helices possess greater conformation diversity than α-helices:
To obtain an exact and general measure of the variability of conformational properties of $3_{10}$ and π-helices, a thorough study of RH torsion angles (φ-ψ) was conducted. Furthermore, to compare the range of conformational variability observed for RHs torsion φ-ψ, examination was performed on φ-ψ angles of stable secondary structure α-helices too. For this analysis, residues from high-resolution protein crystal structures (< 2 Å) were considered. Secondary structural states of these residues were obtained from DSSP [42] (based upon hydrogen-bonding criteria). Ramachandran maps [43] constructed from torsion angles subtended by all non-glycine non-proline residues, suggested, that it might be simplistic and misleading to calculate the "mean" range of φ-ψ angles to describe the conformational properties of RH. Grounds for such assertion stems from the fact RH φ-ψ angles were found to be distributed across all four quadrants of Ramachandran map, even though only non-glycine non-proline residues were considered (**Fig. 1**). Furthermore, many RH torsion angles were found to reside in normally disallowed regions of Ramachandran map.

Although studies on the torsion angles of proteins are not uncommon, results obtained from the present work are unexpected. **Table 1** suit reveals that 98% of non-gly, non-pro α-helix residues tends to populate the 'core' region of Ramachandran map. This is expected from a regular secondary structure, with which definite structural and/or functional roles could be attributed in particular as well as in general terms. However, amongst $3_{10}$-helix non-glycine, non-proline residues, only 88% are found to populate the 'core' region. For the π-helices, merely 82% of non-glycine, non-proline residues could be found in the 'core' region. Normalized percentage of non-glycine, non-proline residues in $3_{10}$, π and α-helices supported these trends from a different point of view. One finds that normalized percentage of all the residues in 'disallowed' region for $3_{10}$, π-helices to be 1.54 and 2.0, which are more than double than the same for α-helix residues, a mere 0.75. Furthermore, it was remarkable to note that, after discarding glycine and proline, normalized percentages of $3_{10}$ and π-helix residues in 'disallowed' region had gone up by 19% (from 1.29 to 1.54) and (staggering) 63% (from 1.23 to 2.00) respectively; whereas, the same for α-helices remained exactly the same. Noteworthy also is the fact that while one can observe significant proportion of $3_{10}$ and π-helix residues as part of 'allowed' regions in **Table 1**; for α-helices, only 2.5% of the residues were found to be residing in the 'allowed' region; which in turn, decreases further to a trifle 1.3% for non-proline and non-glycine residues. Such observation presents a stark contrast between number of α-helix residues in 'core' region and that in 'allowed' region, pointing to an inherent tendency of α-helix residues to be in the best stereochemically favored region. It was equally interesting to note that while there is 70% retention of $3_{10}$-helix residues in the 'disallowed' region, even if glycine and proline are not considered; for the same in α-helices, only 62.63% retention could be observed.

All of these findings seem to imply innate propensities for the $3_{10}$ and π-helix residues in regions outside the 'core' area of Ramachandran map; which in turn, appears to be in sharp contrast to the intrinsic tendency for α-helix residues, which distinctly preferred the 'core' region. This difference suggested that, nature did not employ the same preferential (favored) biases in its organizational principles while constructing the RHs, as it did while constructing the more stable secondary structure viz. α-helix.

**Probabilistic characterization of the occurrence profile of $3_{10}$ and π-helices on protein sequence:**

The identification of RH for comparative conformational flexibility study was based on the hydrogen-bonding pattern. However, the large conformational space allowed to achieve *i→(i+3)* and *i→(i+5)* hydrogen bonding posed another interesting question; namely, to what extent is it possible to identify these RH, or estimate their number in any protein, anchoring only on the torsion angle information? Hollingsworth et al. [38] (based on torsion angle analysis from ultra-high resolution crystal structures) argued that there are only three linear groups in proteins; the extended conformations of β-strands, the cluster of α-and-$3_{10}$ helices, and the broad population of polyproline-II-like spirals. To test various relevant aspects of this argument on a broader scale and also to explore the patterns in the rare occurrences of RH, we developed a probabilistic model based on two-fold observations - from literature and from torsion angle-based investigation described in the preceding section. Rather than predicting the exact coordinates of RH, our model tries to ascertain the number of "mean occurrences" ($m_x$) of $3_{10}$ (and separately) π-helices in a protein backbone; which, in other words, are the expectation values of probability of occurrence of $3_{10}$ and π-helices in any protein structure. This was necessary requirement in order to decipher the global pattern in RH occurrence profile across protein universe – one, that cannot be determined from case-specific attempts to predict individual occurrences of either $3_{10}$ or π-helices.

Predicted results from the model were compared with the "actual" mean occurrences of $3_{10}$-helices for every protein, which were obtained from DSSP. Where the predictions from our model were made by banking upon the torsion angle information of three consecutive residues, DSSP identified the $3_{10}$ or π-helices by measuring hydrogen bond energies. The prediction-and-verification scheme, thereby, could accommodate two complementary approaches. Accuracy of the obtained results vindicated the underlying assumption behind our model; namely, RH occurrence profiles on the primary structures follow stochastic distribution. However, not to miss out on the context-specific constraints on particular primary structures, the globally predicted mean occurrence value was validated for every individual protein by the use of a 'correction parameter', before starting the analysis (please refer to **Materials and Methods** for details of model construction). Thereby, both the concerns, the global general trends and local constraints of particular cases, could be addressed simultaneously in an unified quantitative scheme.

Present model served two purposes; first, it established the stochastic nature of RH occurrence on primary structures, validating thereby the basic assumption behind construction of the model. This, in turn, suggested that sequence-dependent occurrence of $3_{10}$-helices as observed in helical peptides [39-40], should not be generalized to the (complex) of protein sequences. Second, success of the model confirmed the argument put forward by Hollingsworth et al. [38], that, even with extreme conformational diversity one can identify the linear groups in proteins with substantial accuracy (mean occurrence profiles of $3_{10}$ and π-helices in the present case). To obtain the global structural view, analysis was carried out on a protein 'structural class'-specific manner. Protein structural classes were obtained from SCOP classification [44].

Predicted trends were plotted against the actual trends, and could be observed to follow the actual occurrence profile of $3_{10}$-helices in general (**Fig. 2**). However, for almost all the structural classes, these trends portrayed a consistent profile of under-prediction of predicted magnitudes; suggesting a systemic under-prediction due to the algorithm proposed here. Nonetheless, extent of this under-prediction was observed to be small. Maximum margin of error for all-α (**Fig. 2A**) and membrane proteins (**Fig. 2D**) was found to be ~0.1 probability unit, for small proteins (**Fig. 2F**) it was found to be ~0.085 probability unit, for α/β (**Fig. 2B**) it was ~ 0.08 probability unit, and for α+β 0.06 probability unit (**Fig. 2C**). For the multi-domain proteins (**Fig. 2E**), the least error was observed - mere 0.05 probability units. In terms of exact number of errors, these probability values implied that for the entire structural class of all-α proteins, present algorithm will be under-predicting a maximum of ~38 $3_{10}$-helices with probability 0.1; ~23, with probability 0.2, ~24, with probability 0.3 and so on. Similarly, for the multi-domain proteins, a maximum of ~10 $3_{10}$-helices might be under-predicted with probability 0.05; while for small-proteins, ~20 $3_{10}$-helices might be under-predicted with probability 0.08. When applied on π-helices, this algorithm could predict the mean occurrence of π-helices with high efficiency with maximum error margin < 0.025 probability units (**Fig. 3**). This implied that the proposed probabilistic construct could model the global statistical characteristics in the occurrence profile of π-helices better than it could for $3_{10}$-helices.

For the investigation behind small yet definite margin of under-prediction, 30% of the proteins from our dataset were selected randomly, for manual examination. Such (manual) examination revealed that, due to significant variation in the range of torsion angles in $3_{10}$-helices (for all SCOP classes) the filtering criterion for detection of $3_{10}$-helices (based on torsion angles) fell short in identifying them. Huge variations in torsion angle magnitudes were anticipated, especially while working on a large set of $3_{10}$-helices. The 'correction parameter' (details in '**Methods**') was introduced precisely to cope with the highly irregular pattern of $3_{10}$-helix φ-ψ combinations. However, it fell short of capturing the sheer extent of φ-ψ variability, albeit by a small margin. This was, nevertheless, expected as the correct prediction of $3_{10}$-helices is difficult and position specific shifts in the torsion angles of $3_{10}$-helices have also been reported [1, 45]. Hence the present algorithm could not detect ~2% of $3_{10}$-helices in the multi-domain proteins, ~4% $3_{10}$-helices in α+β and membrane proteins, and 6-8% of the residues belonging to all-α and small proteins. This consistent error in the detection of residues belonging to $3_{10}$-helices explained the small yet measurable differences between 'predicted' and 'real' trends in **Fig. 2A-2F**.

The total number of π-helices from the same set of non-redundant proteins was significantly less than the number of $3_{10}$-helices. Hence, the torsion angle-based filtering scheme to detect the π-helices could be applied more accurately. Thus the detection scheme for π-helices did not suffer from errors that were pertinent for $3_{10}$-helix occurrence pattern study. Distribution of π-helices on the primary structure (**Fig. 3**) provided a clear indication that their occurrences on the primary structure take place in a purely random and accidental manner. The maximum error for the present algorithm to detect π-helices was found to be less than 0.03; implying that the present algorithm could reproduce the patterns of occurrence of π-helices in faithful manner. This in turn, implied that occurrences of π-helices could indeed be termed as Poisson events, suggesting unambiguously that occurrences of π-helices are stochastic events. On a different note, the very fact that π-helices could be identified reliably banking solely upon the torsion angle information, tends to oppose the assertion made by Hollingsworth et al. [38]; namely, π-helices in proteins should not be viewed as a true linear group.

The probabilistic model proposed here did not take into account special knowledge-based considerations about the coordinate of RH on the primary structures. Hence, success of the probabilistic model, on a completely different note, suggests that the popular notion [46-47], namely - $3_{10}$ and π-helices can well be considered as common deformations at the end of α–helices in protein crystals – to be simplistic and inadequate.

**Analysis of sequence intervals between consecutive occurrences of $3_{10}$ and π-helices confirms inherent Poisson nature:**
To draw robust inference about the stochastic nature of occurrences of RHs on the primary structures, it was not sufficient to merely study the occurrence profile of RHs. Hence, to complement the investigation of last section, an independent analysis of the patterns in the inter-arrival sequence-intervals was conducted. Such a statistical study of sequence intervals between consecutive occurrences of $3_{10}$ and (separately) π-helices - helped us in further characterizing the nature of RH occurrences. Investigation was performed on protein sequences across different SCOP classes. Subsequently, maximum likelihood fitting of statistical distributions were performed, to identify the distributions that best describe the inter-occurrence pattern of $3_{10}$ and π-helices in different structural classes.

Results presented in **Fig. 4** and **Fig. 5** point again to the Poissonian nature of sequence intervals between consecutive occurrences of $3_{10}$-helices and (separately) π-helices. One can observe the broad consensus from these plots that, for all the structural classes, the intervals between consecutive occurrences of $3_{10}$-helices and (separately) π-helices are best described by Gamma, Exponential and Weibull distribution, which are found to feature regularly in the top three best fitting distributions, irrespective of the structural class concerned. The Gamma distribution arises naturally for intervals between events having Poisson distribution, whereas Exponential distribution (a special case of Gamma) routinely describes lengths of the inter-arrival duration in a homogeneous Poisson process [48-50]. (In fact, Exponential is the only continuous memory-less random distribution). The fact that these two distributions dominated the list of templates in the $\chi^2$ – best-fit test for inter-occurrence intervals for $3_{10}$-helices and π-helices across all the SCOP classes, pointed strongly to the fact that $3_{10}$ and π-helices occur randomly and accidentally on protein primary structures. Furthermore, this showed that the difference between the 'predicted' and 'experimentally observed' trends in **Fig. 2** and **Fig. 3** were indeed due to failure of detection of $3_{10}$-helices in the primary structures purely, and not due to any erroneous and/or simplistic aspect in the basic hypothesis of the current work.

However, presence of Weibull distribution as one of the best fit template for α+β, multidomain and small proteins indicates, that the "random and accidental" (i.e. Poisson-type) categorization might not be appropriate to characterize occurrences of $3_{10}$ and $\pi$-helices in cases of the these structural classes. This introduces a cautionary note to the inference but does not influence the general deduction about stochastic, rare and accidental nature of occurrences of $3_{10}$ and $\pi$-helices across other structural classes.

The fact that occurrence profile of RHs on the primary structures cannot always be definitely classified as purely stochastic, however, leads us to the more fundamental question; namely, why, for certain structural classes, the occurrence of $3_{10}$ and $\pi$-helices is completely stochastic, whereas for some others the stochastic nature is partial? We chose to investigate this question from the evolutionary perspective.

**Analysis of evolution of rare secondary structures: $3_{10}$ and $\pi$-helices**
Protein structures evolve continuously over time. Secondary structures play an important role in this molecular evolution, as they are an integral part of protein structure hierarchy, connecting the primary structures to biologically functional form. However, protein evolution has mostly been studied from the perspective of considering either the entire sequence or entire structure as a whole. Though one finds some pioneering works [51-55], the role of secondary structures in evolution is far from being exhaustively probed. For example, even though a recent work [45] has pointed out the severely non-conserved nature of $3_{10}$-helices, it could not throw light on the nature of evolutionary transformations in the paradigm of secondary structures. We investigated trends in transformation of $3_{10}$ and $\pi$-helices during long-term evolution of protein structures and possible implications of their stochastic occurrences (as discussed in foregoing sections).

To study the long-term evolutionary changes on these RH we started from protein families that represent possible homology among sequences within a family. We conducted our analysis on twenty largest protein sequence families (drawn from Pfam (Protein Family Database, release 24.0) [56]) under the assumption that all the sequences in these families are descendants of one (or a few) ancestral sequence(s) in the last common ancestor of all the organisms making up the family. From these families, we obtained the proteins with known structures and devised an algorithm to study the cases of *replacement* of one secondary structure with another (***Rep***), *retention* of the same secondary structure (***Ret***), and *insertion* of a new secondary structure (***Ins***) on protein sequence axes. Our studies were confined to the level of secondary structures; we did not resort to analyze of tertiary structures. While describing protein evolution from the perspective of individual secondary structural elements, $3_{10}$ and $\pi$-helices were considered as the units of these changes, just like an insertion or substitution of a single amino acid in sequence. Although it might not be plausible to assume that a single mutation of one residue in a secondary structure will always account for the change of the entire secondary structure, especially within a short time-scale; in case of cumulative mutations taking place over long-term evolutionary timescale, such change of an entire secondary structure can be expected. Such an approach appeared to be logical owing to the fact that Pfam families under study contains proteins from a wide verity of organisms ranging from bacteria to humans; and the structural data was extremely sparse.

Results (details are provided in **Fig. 6A** and in **Supplementary Material**), expectedly, indicated that in six out of the twelve families (PF00115, PF00072, PF02518, PF00069, PF00106,

PF00583), number of events of retention (***Ret***) of $3_{10}$-helices were found to be less than both the number of events of their replacement (***Rep***) and number of events of their new insertion (***Ins***), independently. The largest difference between (***Rep***) and (***Ret***) (|***Rep***| - |***Ret***| = 1864) could be noted for PF00069, viz. the protein kinase domain; while large differences (|***Rep***| - |***Ret***| ~ 500) could be observed for both PF00106 (short chain dehydrogenase) and PF00583 (Acetyltransferase family). Only in two Pfam families, (***Ret***) was found to be greater than (***Rep***), but this finding may not be provided with much importance, because number of $3_{10}$-helices in both of these families were statistically insignificant. For example, in PF00078 (the reverse transcriptase) ten $3_{10}$-helices were retained while six were found to be replaced; whereas in PF00528 (Binding-protein-dependent transport system inner membrane component) only two $3_{10}$-helices were retained, while one was found to be replaced. For three (PF00005, PF00033, PF00032) out of twelve families, (***Rep***) outnumbered both (***Ret***) and (***Ins***), independently. Perhaps most importantly, (|***Rep***|>|***Ret***|) was observed in nine out of twelve cases. This appeared to be the most significant trend, because other (three) cases with (|***Rep***|<|***Ret***|) trend were found to be characterized by statistically insignificant number of $3_{10}$-helices. The last observation, viz. $3_{10}$-helix replacements outnumbering $3_{10}$-helix retention in 75% of the cases, takes the non-retention trend to a logical finish. One therefore notices a global trend of non-retention of $3_{10}$-helices, which went on to support the general hypothesis behind the work. There was just one class (PF00005) having 15 π-helices, all of them were found to be replaced by α–helices.

The evolutionary nature of non-retention of $3_{10}$-helices could be ascertained more clearly from **Fig. 7** where frequencies of events of (***Rep***)*,* (***Ret***) and (***Ins***) of $3_{10}$-helices have been plotted over increasing evolutionary distances. It can be seen that at low evolutionary distances between two proteins within a family, events of (***Ret***) and (***Ins***) are larger, while as the proteins' distance increase further, events of (***Rep***) outnumber both (***Ret***) and (***Ins***)*.* Furthermore, events of (***Ret***) is significantly lesser than both new (***Ins***) and (***Rep***)*.*

However, certain unexpected results were also observed. For example, in case of Binding-protein-dependent transport system inner membrane component class (PF00528). Proteins of this Pfam family are multi-component systems typically composed of a periplasmic substrate-binding protein, one or two reciprocally homologous integral inner-membrane proteins and one or two peripheral membrane ATP-binding proteins that couple energy to the active transport system. Apart from a conserved region at the C-terminal extremity, sequence of these proteins is quite divergent, and they have a variable number of trans-membrane helices. Almost 70% of $3_{10}$-helices are retained for Proteins of PF00528; 30% of them undergo replacements with equal probability (50%) to α-helices and 'turns'. However, absolutely no instance of insertion of $3_{10}$-helices could be found in them.

This observation was remarkable because one can explain such data from two contradictory logical grounds. One, due to evolutionary pressure, insertion of new $3_{10}$-helices is prohibited in PF00528 proteins; and furthermore, PF00528 proteins are replacing all their $3_{10}$-helices to other secondary structures, with whom definite importance (either structurally or functionally, or both) can be associated. But this reasoning fails to account for only 30% of replacement of $3_{10}$-helices, and not more. A second logic, from just the opposite perspective, might argue that for PF00528 proteins, $3_{10}$-helices do indeed entail some structural and/or functional importance. Retainment of 70% of $3_{10}$-helices from a pool of divergent sequences, appears to support this argument.

However, this view fails to interpret why at all replacement of $3_{10}$-helix will take place and why, insertion of new $3_{10}$-helices will be prohibited if they indeed are structurally and/or functionally important for PF00528 proteins. Only a thorough structural survey of $3_{10}$-helices in PF00528 proteins at a global level, can resolve this debate. However, such an analysis is outside the scope of the present study.

However, one notices the trend ($|\mathbf{\textit{Ins}}|>|\mathbf{\textit{Rep}}|$) that in seven out of twelve classes (PF00078, PF00115, PF00072, PF02518, PF00069, PF00106, PF00583). Most staggering case of insertion of $3_{10}$-helices over their replacements, could be observed in protein kinase domains, ($|\mathbf{\textit{Ins}} - \mathbf{\textit{Rep}}|$ = 6517); whereas in PF00106 (short chain dehydrogenase) and PF00583 (Acetyltransferase (GNAT) family), large differences ($|\mathbf{\textit{Ins}} - \mathbf{\textit{Rep}}| > 500$) could be observed. These findings tend to imply that, for proteins belonging to certain Pfam classes, $3_{10}$-helices are not being replaced by other secondary structures; rather, new $3_{10}$-helices are being introduced continuously. Quite unmistakably, this is in contradiction to our hypothesis that $3_{10}$ and π-helices are mere accidentally occurring structures, with little importance to proteins. Although, comparable number of events of $3_{10}$-helix insertions and replacements ($|\mathbf{\textit{Ins}} \sim \mathbf{\textit{Rep}}|$) emerged as the predominant trend in four out of seven Pfam classes, broad trend of the results pointed at significant number of $3_{10}$-helix insertions too. Reasons and implications of these findings are discussed in detail later.

**Non-conservation of $3_{10}$ and π-helices over long-term evolutionary process, transformation of $3_{10}$-helices to stable and/or functional secondary structures:**
Next, we focused on the observed replacement events. Looking at the secondary structures which replace the RH, we observed that they are primarily replaced by more stable secondary structures. Thus, as results presented in **Fig 6.B** depicts, consistent trends of $3_{10}$-helix→α-helix transformations, or $3_{10}$-helix→loop transformations, or $3_{10}$-helix→β-sheet transformations, could be observed; together with equally ever-present trend of $3_{10}$-helix insertion. The number of $3_{10}$-helix→α-helix transformations was found to be more than the number of $3_{10}$-helix→loop transformations, in six out of eleven cases; in three out of eleven cases this trend reversed; while in two out of eleven cases $3_{10}$-helices were found to show equal tendency to be transformed to either α-helices or to loops. Although some instances of nontrivial number of $3_{10}$-helix→β-sheet transformations could be observed, in case of no Pfam class, the events of $3_{10}$-helix→β-sheet transformations could outnumber either of $3_{10}$-helix→α-helix or $3_{10}$-helix→loop transformation trends.

In order to release the steric constraints, it is easier for a $3_{10}$-helix to transform itself to either α-helix or a loop, than to undergo a large-scale rearrangement to achieve a state of local energy minima in the form of β-sheet. This explains the reason behind low instances of $3_{10}$-helix → β-sheet transformation, in comparison to instances of $3_{10}$-helix→α-helix, or $3_{10}$-helix→loop transformations. Very significantly, in no single case $3_{10}$-helix→π-helix transformation or π-helix→$3_{10}$-helix transformation was observed during the course of long-term evolution.

**Relationship between SCOP classes and trends in retention, replacements and insertions of the $3_{10}$ and π-helices:**

No correlation could be found between SCOP structural classes and trends in the occurrence of anyone of **Rep**, **Ret** and **Ins**. For example, while for many PF00005 proteins (ABC transporters) with structural domain α/β the trend (***Rep*>*Ret*>*Ins***) was observed; for PF00106 (short chain dehydrogenase) with the same structural domain α/β, the trend changed to (***Ret*<*Rep*<*Ins***). On the other hand, although the ratio **Rep : Ret : Ins** is found to be almost identical for PF00106 (short chain dehydrogenase) and PF00115 (Cytochrome C and Quinol oxidase polypeptide I), SCOP domains of proteins associated with these families are found to be vastly different. SCOP-specific pattern search for the three cases where ***Rep*** outnumbered ***Ret*** and ***Ins*** both (viz. PF00005: ABC transporters, PF00033: Cytochrome-b (N-terminal)/b6/petB, and PF00032: Cytochrome b(C-terminal)/b6/petD); failed to reveal any clear correlation.

## Discussion:

The present study attempted to investigate the scopes of some fundamental questions, viz. what are the patterns that describe the occurrences of RHs on primary structures? What are the structural and/or functional reasons behind the existence of RHs - in general? What are the reasons behind contradictory findings about RHs? If RHs are at all crucial for structural and/or functional integrity of proteins; what kind of constrains do they impose on protein structures over a long period of time? Finally how do their presence shape the evolutionary trajectory of proteins at structural level?

We could demonstrate that occurrences of both $3_{10}$ and π-helices on the primary structures can be considered as Poisson events, to a large extent. While this Poissonian trend is extremely evident for proteins belonging to certain structural classes, for some others, the extent of Poisson nature is comparatively less. Nevertheless, both occurrence profiles as well as inter-arrival intervals tend to provide unambiguous evidence for Poisson nature in the distribution of RHs on primary structures. Since the present study involved the entire universe of available non-redundant protein structures, we argue that the aforementioned assertion can be considered general.

Properties of RHs are in sharp contrast with that of α-helices. For example, it is known that RHs are geometrically irregular, inherently unstable and typically small, ranging from 3 to 4 residues on an average [30-31]. Residues forming RHs are less than 3% of the total number of protein residues even with a lenient criterion [5, 31, 57]. In contrast, α-helices are relatively regular and longer secondary structures, accounting for 32-38% of all protein residues [42, 58], and frequently composed of more than 15 residues. Our evolutionary findings quantified the trend of non-retention of RHs clearly. Most remarkably not a single case could be observed where the event $3_{10}$-helix→π-helix transformation or π-helix→$3_{10}$-helix transformation had taken place. Such an observation rules out any possibility of inter-conversion of one unstable helix into another. Rather, as general and consistent trends, $3_{10}$-helix→α-helix, π-helix→α-helix and $3_{10}$-helix→'loop' transformations were observed. In certain cases $3_{10}$-helix→β-sheet transformations were observed too. Since α-helix, β-sheet, loops – are all known to be stable secondary structures, with which definite structural and/or functional importance are attached regularly; one might interpret these trends as evolutionary mechanisms to achieve optimal structural and/or functional benefits that could not be achieved with either $3_{10}$ or π-helices.

We propose that the aforementioned observations and the apparently disparate array of results regarding various known facts about RHs (as mentioned in the **Introduction**) can be explained from a single unifying perspective, if it is hypothesized that $3_{10}$ and π-helices symbolize 'imperfect structures' formed during the protein folding process. Since these imperfections take place rarely, one finds appropriate explanation behind rare occurrences and much smaller lengths of $3_{10}$ and π-helices. This explains the reason behind direct proportionality between length and irregularity-instability of $3_{10}$-helices, as reported in a recent work [1]. Such rationalization, on one hand, explains why less than 3% of the total population of PDB-residues reside in RH secondary structural states; while, on the other hand, it explains why most $3_{10}$-helices are composed of only 3 residues and also, why one rarely finds any $3_{10}$-helix with more than 5 residues. The same rationalization explains why, even though the first $3_{10}$ helical turn leads to less entropic loss than first α-helical turn [59] (it requires 3 instead of 4 residues need to be constrained), occurrences of $3_{10}$-helices are comparatively rare.

Although fluctuations in torsion angles of protein residues are ubiquitous for each secondary structure [60, 61]; yet, as revealed by **Fig. 1** and **Table 1**, for both $3_{10}$ and π-helices the aforementioned deviations of φ-ψ values is much more pervasive than what could be observed for φ-ψ values of α-helices. Such smeared φ-ψ profile explains why the RH geometries tend to be irregular, as compared to the case of α-helices. These RHs are formed during folding process, possibly, just to accommodate the residues in other favorable secondary structural states and not due to absolute requirements of [i →(i + 3)] and/or [i →(i + 5)] hydrogen bonding pattern. The unambiguous Poisson characteristics observed in the occurrence profile RHs emphasize the possible ad-hoc status of $3_{10}$ and π-helices in protein structures. If there had been an exact requirement for such an hydrogen bonding pattern, their occurrences could not have been accidental and rare (hallmark of Poisson distribution). Instead, if it were hypothesized that these RHs are just the rarely formed imperfectly folded structures, the greater torsion angle diversity of theirs compared to the regular secondary structures – could be explained neatly. Indeed, despite sharing the same energy minima with the α-helices in the (φ-ψ) space [26] and possessing an α-helix-like CD spectrum [62], one observes considerable extent of (φ,ψ) angle variation in $3_{10}$-helices. Variation in the (φ-ψ) magnitude in α-helices can be easily be detected too; however, such variations can always be observed to be limited within a small area of Ramachandran map, implying clearly a categorical plan of nature to assign definite significance to [i →(i + 4)] hydrogen bonding pattern, and not to [i →(i + 3)] and [i →(i + 5)] hydrogen bonding. Possibly the same logic can be extended to other incongruous reports about role of RH in the paradigm of helix-coil transition and helix formation in general.

Extensive smearing of RH φ-ψ values can be interpreted from the other way round too. Since the φ-ψ deviation range for the regular secondary structures are less, one can interpret that less deviation in φ-ψ values as a necessary pre-requisite for regular secondary structures to acquire significant structural and/or functional roles. Inconsistent φ-ψ profiles for the regular secondary structures, probably, would have accounted for less reliability in their ensuring definite structural and/or functional responsibilities. However, since wide-ranging variations in φ-ψ magnitudes were observed for $3_{10}$ and π-helices, one can consider this to be a prime hindrance against their acquiring definitive structural and/or functional roles. Drawing from the previously stated hypothesis, one can reason that the absence of indispensable structural and/or functional roles of RH might well be due to extensive variations in their φ-ψ magnitudes. (One can note the extent of such non-indispensability of RHs by observing that - no study could be found to document a statistically significant number of cases; even, as low as five cases, where $3_{10}$ or π-helices were found to perform certain structural and/or functional roles that α-helices in the same positions would have failed to perform.) Since nature viewed the $3_{10}$ and π-helices as aborted structures, hardly ever any indispensable structural and/or functional role was assigned to them. This explains why, functional associations of RHs are negligible.

We also find the support of this logic in the evolutionary study where it is evident that the non-retention of $3_{10}$-helices is a prominent trend in most of the Pfam classes under study. Such a trend is even more pronounced in highly conserved proteins, for example, in protein kinases (PF00069) domain. It is well known that function of protein kinases is evolutionarily conserved from E. coli to human; and furthermore, catalytic subunits of protein kinases are highly conserved. One finds (**Fig. 6A**) that retention of $3_{10}$-helices in protein kinases is the minimum amongst the top 20 Pfam protein families considered in the present study (***Rep***$_{PF00069}$ = 2714, ***Ret***$_{PF00069}$ = 850, ***Ins***$_{PF00069}$ = 9231). This suggests that to ensure the evolutionarily conserved nature of their structure and function, the disadvantageous structures like RHs were not conserved and were quickly replaced by other, more stable, secondary structures.

**$3_{10}$ and π-helices as evolutionary intermediates of α-helices**
Based the hypothesis that $3_{10}$ and π-helices are imperfect structures, one might argue that mutations that tend to convert $3_{10}$ and π-helices into more stable structure like α- helices, will be favored by evolution. However, evolutionary transformation study (presented in '**Results**' section) revealed many cases where the number of new $3_{10}$-helix formation, viz. (***Ins***), exceeded the number of cases of their replacements (***Rep***). This is contrary to our hypothesis. In addition, such observation of insertion of new $3_{10}$-helices appears to contradict the views expressed in another previous work, where $3_{10}$-helices were categorized as "para-helices" [1]. Thus, this particular observation presents an interesting problem in the realm of evolutionary dynamics of proteins at the secondary structural level.

But this apparent contradiction can be resolved by remembering that evolution is a continuous process, which does not proceed at a constant rate [43]. Furthermore, during evolution, many $3_{10}$-helices can be inducted into structures by random mutations, only if these events turn out to be largely neutral and not deleterious. Hence, while many $3_{10}$-helices are being constantly inducted into the process of protein evolution, many of them, subsequently, are being replaced to more stable secondary structures also. Therefore, ubiquitous trends of $3_{10}$-helix→α-helix transformations, or $3_{10}$-helix→turn transformations, or $3_{10}$-helix→β-sheet transformations, could be observed; together with equally ever-present trend of $3_{10}$-helix insertion. Furthermore, it appears that under continuous process of mutation and selection at molecular level, $3_{10}$ and π-helices can best be viewed as evolutionary intermediates of other more stable secondary structures.

The suggestion that $3_{10}$-helices can be considered as an intermediate in the folding/unfolding of α-helices was made previously too [20]. However, such a suggestion was not made from the perspective of evolution of secondary structures, but was based on the principle that there is a lower entropic penalty for the necessary loop closure for the formation of i→i+3, as compared to i→i+4 hydrogen bonds. Furthermore, there is no disallowed region of Ramachandran map between the preferred φ-ψ magnitude for $3_{10}$ and α-helices [63] and, the $3_{10}$ block of φ-ψ is energetically less favorable [7, 32]. This smart quality control mechanism in potential energy space helps in ensuring a smooth $3_{10}$-helix→α-helix transition, but not an α-helix→$3_{10}$-helix transition. Similar argument can be proposed for the π-helices too; as it has been recently reported that π-helix block of φ-ψ is energetically less favorable than that of α-helices [64]. The present assertion from evolutionary analysis, viz. $3_{10}$-helices are unrealized possibilities in their route to become α-helices, find support from the findings of many recent studies too, where $3_{10}$ and π-helices are typified as "transient" and "defective" α-helices [65-68]. The current study could therefore provide a unified reasoning (both structural and statistical) to previous assertions, forwarding alongside a possible, deeper, cause to describe every aspect of statistical findings about RHs, from structural perspective.

The facts that - φ-ψ zone between $3_{10}$ and α-helices is energetically continuous; a $3_{10}$-helix→α-helix transition is energetically favorable but an α-helix→$3_{10}$-helix transformation is (highly) unfavorable; and evolutionary trends of insertions and replacements of $3_{10}$-helices (by α-helices) are ubiquitous – can be brought together in one platform in an attempt to understand why many $3_{10}$-helices are found either at the beginning or at the end of α-helices. While the process of induction of a $3_{10}$-helix might be evolutionary, due to the inherent problems with i→i+3 hydrogen bonds, they are quickly replaced by favorable i→i+4 bonds. Such a scheme of an imperfect structure ($3_{10}$-helix) being a structural evolutionary intermediate for a stable structure (α-helix) – can neatly explain $3_{10}$-helices preceding α-helices, but it cannot explain the non-trivial number of cases where $3_{10}$-helices are found at the end of α-helices. Whether these occurrences can be considered as abrupt ends of folding of concerned α-helices, or whether these occurrences merely depict the snapshots of evolutionary process where the concerned $3_{10}$-helices are on their route to transform themselves to either α-helices or turns – could not be established from the sphere of the present investigation. Only an exhaustive and dedicated study that takes into account all of the aforementioned facts and patterns and that probes the $3_{10}$-helices succeeding α-helices – can decipher the evolutionary and structural reasons for their presence.

$3_{10}$-helix→turn transformations are observed in significant number of cases too. The fact that torsion angle range of type III turns, ($-60^0$, $-30^0$), matches closely with that of many $3_{10}$-helices, suggests that viewing $3_{10}$-helices as evolutionary intermediate for turns – might not be incorrect either. This interpretation receives further support from the observation of a previous report that a three-residue $3_{10}$-helix may as well be viewed as two overlapping β-turns. With the evolutionary outlook presented in the current work, one can explain why, the N3 position of $3_{10}$-helices and i+2 positions of β-turns share similar (φ, ψ) distributions and similar residue propensities, as was reported in the aforementioned work [10].

Hence, to summarize, alongside the replacement of existing $3_{10}$ and π-helices, new $3_{10}$ and π-helices are being introduced into proteins too, by continuous mutation exposure. In due course, a fraction of them undergo (likely) replacement, while the rest are retained for longer, before being replaced by either α-helices or turns at that structural location. This finely balanced interplay of continuous processes underscores the dynamic equilibrium persisting in protein structures under structural, functional and evolutionary constraints.

## Materials and Methods:

**Protein Structure Data collection and Secondary Structure Identification**
Protein crystal structure information was collected from Protein Data Bank [69]. In order to remove redundancy, only structures having resolution better than 3Å and less than 70% sequence identity were considered. To ensure the most general and exhaustive nature of the dataset, all protein structures with the two aforementioned criteria, were considered. Torsion angles and secondary structure annotations for these structures were obtained from DSSP [42]. The helices identified as $3_{10}$ and π-helices by DSSP algorithm were considered to be 'actual', whereas the ones identified by the present algorithm (using torsion angle range alone) - were considered as 'predicted' helices. (Choice of DSSP as the software was not arbitrary. Due to possible incompatibility with certain experimental measurements in the mode of identifying the $3_{10}$-helices [67], STRIDE [70] was not chosen for the purpose.)

The non-redundant structures were divided into different structural classes as per SCOP definition [44]. Only the 'true' classes from SCOP were taken into account (viz. all-α, all-β, α+β, α/β, membrane, multi-domain and small proteins). Investigations were performed in a SCOP class-specific manner to identify variability due to differences in the distribution of $3_{10}$ and π-helices, as specific to different structural classes. Since the number of π-helices is extremely small and is statistically insignificant for any single SCOP class, structural class-specific analysis could not be performed for π-helices. Thus the entire set of π-helices was considered as a single unit. Although a previous work [5] had reported the presence of $3_{10}$-helices in all-β proteins, DSSP could not assign any $3_{10}$ or π-helix in all-β protein class; therefore this class was ignored from analysis.

**Ramachandran Plot construction:**
For Ramachandran plot study, only the protein structures with very high resolution (less than 2 Å) were considered. For each structure, the torsion angles φ and ψ were taken from DSSP. Ramachandran plot was constructed from aforementioned non-redundant dataset of protein structures for all structural classes. The definition of core, generous, allowed and disallowed regions were taken from PROCHECK [71].

**Modeling the Occurrence Profile of RHs**
Set of careful observations regarding nature of RH occurrence on protein sequence, formed the foundation for methods applied for model construction. While observations of biophysical nature have already been talked about in the 'Discussion' section, it is important, to discuss the mathematical ones too, before delving into the methodological details.

An appropriate distribution that attempts to describe the occurrence profile of the $3_{10}$-helices should be continuous in nature. (This is because the 'random variable' attempting to describe the aforementioned occurrence profile should be capable of assuming a continuous range of values, instead of fixed and discreet values.) While scanning through an arbitrarily chosen primary structure, an observer may notice a $3_{10}$-helix at coordinates $S_1$, $S_2$,… $S_i$; where $S_i$ denotes the $i^{th}$ occurrence of $3_{10}$-helix on a primary structure (S). The required distribution should be able to predict the sequence length that one needs to scan before detecting another $3_{10}$-helix on the primary structure. We may now observe the following characteristics of $3_{10}$-helix occurrence.

**1.** No periodic (or other global) patterns could be detected in the occurrence profile of $3_{10}$-helices. Indeed no such report could be found from existing literature too.
**2.** Two or more $3_{10}$-helices occurring at the same coordinate of primary structure is a physical impossibility. This implies that the occurrence profile of the $3_{10}$-helices constitutes an ordinary flow of events.
**3.** The probability of not finding a $3_{10}$-helix for j units of sequence during a search is the same as that of a fresh search that fails to find a $3_{10}$-helix in the next j unit of sequence. In other words, past history of searching (that is, search conducted along the sequence till the point the search has reached), has no effect on finding or not-finding a $3_{10}$-helix.

Here we note that, all the arguments presented in the context of $3_{10}$-helix occurrence profile, can be extended to π-helix occurrence profile too.

Mathematically, $P(S > s_j + d \mid S > s_j) = P(S > s_j)$ ………… (1)
Where,
P(S): is general probability of detection of a $3_{10}$-helix on the primary structure S,
$j$ : an arbitrary positive variable, relating to the absolute length of sequence interval $|j| < |S|$,
$s_j$: coordinate of primary structure S, describing the position of control of search operation at any given instance.
and
d: distance measured on primary structure S.
- This property suggests that occurrence of $3_{10}$-helices on the primary structure possess the 'memorylessness property' (alternatively, 'evolution without aftereffects') [72-73].

**4.** Differences in the occurrence coordinates of $3_{10}$-helices on the primary structure axis ($S_2$-$S_1$, $S_3$-$S_2$, …, $S_{i+1}$-$S_i$) is stochastically independent for any integer 'i' (i=1,2,…n). Thus, if the event of detection of a $3_{10}$-helix is represented with $D_{Si}$, then for detection events at monotonically increasing primary structure coordinates $S_0$, $S_1$, …, $S_i$ ($0<S_0<S_1< …< S_i$), one can assert that ($D_{S1}$-$D_{S0}$), ($D_{S2}$-$D_{S1}$),…, ($D_{si}$-$D_{si-1}$) are independent for any i.

With these about occurrence profile of $3_{10}$-helices in mind, we can assume that, there exists a real parameter λ (interpretation of which is discussed later in depth) such that, for every 'i' on the distribution of $S_i$, it holds true. As a whole, λ will characterize a Poisson distribution with parameter λ$s$, where s(s⊆S) denotes the length of a randomly chosen continuous stretch of primary structure axis. Interpretation of 's' is provided later.

Since, first, occurrences of $3_{10}$-helices are rare events; second, such occurrences take place independently yet continuously at a constant average rate on any primary structure; Poisson nature of occurrence profile of $3_{10}$-helices becomes apparent. Furthermore, since sequence intervals between two consecutive $3_{10}$-helices are similar to inter-arrival times of events in a homogeneous Poisson process, one can reliably attempt to model the aforementioned intervals on primary structure by exponential distribution [48-50].

To verify the intuitive assessment of the previous paragraph with a rigorous framework, we attempted to construct a formal algorithm that scans through the primary structure, in order to detect and count the $3_{10}$-helices on any given primary structure. Here, search for $3_{10}$-helices take place on a random length L on primary structure, S. The random length can either be the entire length of primary structure under consideration, or a segment of it; whence $L \leq |S|$. Searching operation has an exponential distribution with parameter $\mu(\mu=1/\gamma)$, where $\gamma$ denotes the 'actual' (and not 'predicted') number of occurrence of the $3_{10}$-helices (identified by DSSP) in L. On the basis of aforementioned observations, model was built by assuming that occurrence of $3_{10}$-helices follow a Poisson flow with intensity $\lambda$ ($\lambda$ being the average number of detection of $3_{10}$-helix per unit traversal of L; which implies $\lambda=\lambda(s)$). Finally, s ($s \subseteq S$ and $|s|<|L|$), talked about earlier, was used to denote a subset of L. The parameter s represented the length of the particular randomly-chosen continuous stretch of L, that is being investigated.

Designing a full-proof detection scheme to identify $3_{10}$-helices is extremely difficult. Hence, we introduced the 'correction parameter' $p$ that quantified the probability with which the algorithm could detect and count $3_{10}$-helices. With the introduction of 'correction parameter', global (statistical) trends in occurrence profile of $3_{10}$-helices could be linked to the local constraints of particular cases. Finally, we designated a random variable X, to describe the number of recorded $3_{10}$-helices from a given primary structure. In this work, we derived the distribution and corresponding characteristics of the mean ($m_X$) and variance ($Var_X$) of random variable X, before verifying these theoretical predictions with actual profile of occurrence found from DSSP.

Banking on the aforementioned background, the basic scheme for conditional probability calculation could be set up as:

$$P\{X=\gamma|s\} = \frac{(\lambda ps)^\gamma}{\gamma!} e^{-\lambda ps}$$

Hence, the total probability of the event $\{X=\gamma\}$ is given by:

$$\begin{aligned} P\{X=\gamma|s\} &= \int_0^\infty \left( \frac{(\lambda ps)^\gamma}{\gamma!} e^{-\lambda ps} \mu e^{-\mu s} ds \right) \\ &= \frac{\mu}{\gamma!}(\lambda p)^\gamma \int_0^\infty \left( s^\gamma e^{-(\lambda p+\mu)s} \right) ds \\ &= \mu \frac{(\lambda p)^\gamma}{(\lambda p+\mu)^{\gamma+1}} \end{aligned}$$

$$= \frac{\mu}{(\lambda p+\mu)}\left(\frac{\lambda p}{\lambda p+\mu}\right)^{\gamma} (\gamma=0,1,2,...).$$

This is a geometric distribution with parameter $\frac{\mu}{(\lambda p+\mu)}$, and therefore, the mean and variance of this distribution is given by:

$$\gamma_x = \frac{\left(\frac{\lambda p}{\lambda p+\mu}\right)}{\left(\frac{\mu}{(\lambda p+\mu)}\right)} = \frac{\lambda p}{\mu} \quad \ldots\ldots\ldots\ldots (2)$$

and,

$$Var_x = \frac{\left(\frac{\lambda p}{\lambda p+\mu}\right)}{\left(\frac{\mu}{\lambda p+\mu}\right)^2} = \frac{\lambda p(\lambda p+\mu)}{\mu^2} = \left(\frac{\lambda p}{\mu}\right)^2 + \frac{\lambda p}{\mu} = \gamma_x(\gamma_x+1) \quad \ldots\ldots (3)$$

**Implementation and Verification of the model:**
Accuracy of verification of eq$^n$-2 and eq$^n$-3 with algorithmic constructs rested principally on the reliable detection of $3_{10}$ and $\pi$-helices as (independent) linear group(s), as defined elsewhere[38]. For this purpose, a relaxed torsion angle range was considered that could suitably cover most RH of the dataset. Due to immense diversity in torsion angle range of RH, a fairly large range $\varphi$:($-40^0$ to $-90^0$) and $\psi$:($-25^0$ to $-55^0$) was considered. To reduce the number of false-positives, especially on such a large dataset, only runs of at least three consecutive residues in the aforementioned torsion angle range were considered. The rationale behind this was found from a recent study [38] that established the 3-residue stretch as the shortest yet optimal window to study linear groups in proteins based on torsion angles. Apart from that, the 3-residue stretch served as a reasonable filtering criterion since most RHs are of length three, or more residues.

Denoting the actual number of occurrence any of $3_{10}$-helices in arbitrarily long s(s$\subseteq$S) as $\gamma$; and by defining $\mu=1/\gamma$; $\mu$ and $\gamma$ were calculated. Since occurrences of $3_{10}$-helices are rare and highly non-uniform, considering the entire sequence S in one single attempt to detect the pattern in occurrence of $3_{10}$-helices, may have introduced errors due to coarse-graining. Hence S was sub-divided into overlapping shorter sequences 's', where each 's' was statistically significant in length. Denoting |S|=L and statistically significant length of 's' as 'r' (|r|$\geq$32), the overlapping segments were constructed as [(i=0)$\rightarrow$r; (i=i+1)$\rightarrow$r+1;...; (i=L-r+1)$\rightarrow$L]. With such a scheme the entire sequence was scanned.

This scheme of segmentation was advantageous, because it could take into consideration the (possible) local biases that might have been present in $3_{10}$-helix occurrence in small stretches of primary structure. Calculations were repeated with various magnitudes of r, to identify the aforementioned latent bias. To not miss out on the global picture, $\lambda = (|\gamma|/|S|)$ was calculated too, where $\lambda$ denoted the average number of detection of $3_{10}$-helices per unit traversal of sequence length using torsion angle based assignment of $3_{10}$-helices.

The actual number of $3_{10}$-helices for every protein was obtained from DSSP; this was denoted by $\gamma_x$ (DSSP). $\gamma_x$(DSSP) was equated to $p*(\lambda/\mu)$, for the purpose of finding the magnitude of correction parameter p for every protein. (Correction parameter was necessary to address the fact that certain RH might be missed, even with the best of the efforts to identify them with relaxed torsion angle range. Furthermore, some $3_{10}$-helices may be incorrectly assigned using torsion angle range too). Using the magnitude of p obtained in the last step, the parameter $\gamma_x$(torsion) was calculated for the test cases applying the formula $p*(\lambda/\mu)$. Finally, the magnitudes of means of $\gamma_x$(torsion) and $\gamma_x$(DSSP) were compared, over all the non-redundant proteins, in SCOP-class specific manner, to test the efficiency of probabilistic prediction. The **Fig. 2** suit and **Fig. 3** describes the results where occurrence of particular mean magnitudes of $\gamma_x$(torsion) and $\gamma_x$(DSSP) are plotted (Y-axis) versus number of cases with these mean magnitudes (X-axis). Since $\gamma_x$(torsion) depended pivotally upon $eq^n$-2, this comparison could connect a formal mathematical structure to certain observed traits of secondary structural features within proteins, for which no general quantifiable pattern was suggested hitherto.

It is important to note that the present scheme did not try to probe/predict the exact coordinates of RH on sequences; rather success of the model establishes RHs as randomly distributed linear groups in proteins. The exact coordinates of RHs on sequence may differ from torsion angle identification and DSSP assignment. Therefore we have delimited the scope of our calculations to predictions made with mean occurrence, which can be improved using correction parameter. These mean occurrences were sorted over entire dataset and plotted against actual occurrences in **Fig. 2** and **Fig. 3**. In these figures, the difference in abscissa for any particular ordinate quantified the absolute error (|observed-predicted|) that is occurring with a particular probability (ordinate value). Thus, for any particular magnitude of ordinate, the product of it with the absolute difference between observed and predicted values quantified the number of cases where an error was committed with that particular ordinate probability. For example, for **Fig. 2A** (all-α proteins), for ordinate = 0.1, (absolute) difference between observed and predicted magnitude is ~(960-560) = 400. Hence the product (400 * 0.1) = 40, - quantifies the number of errors made with probability 0.1, in prediction of $3_{10}$-helices in all-α class of proteins. Similarly, one finds 32 [=(~(960-800) * 0.2)] errors with probability 0.2 and 9 [=(~(960-930) * 0.3)] errors with probability 0.3 for all-α proteins. However, such calculations lose meaning for most part of the graphs **Fig. 2C-2F**. This is because the error in prediction is often found to be extremely small, so much so that it is found to involve second decimal place of probability. As a result of such small error values, aforementioned calculation is often found to yield fractional values less than unity. In such a case, for a magnitude of product more than unity, the whole part can be considered as the approximate error of the proposed methodology, for the corresponding probability. For example, for **Fig. 2D** (membrane proteins), one finds error 1.02 [=(~(42-25) * 0.06)] with probability 0.06. This will imply that the present algorithm will commit 1 error with probability 0.06 for membrane proteins.

**Statistical trends in inter-arrival intervals of $3_{10}$ and π-helix occurrences:**
To decipher the patterns of sequence intervals between consecutive occurrences of RH, we calculated the inter-arrival sequence-distances in the observed occurrences of consecutive $3_{10}$ and π-helices (separately). RH definitions were taken from DSSP and calculations were carried out in a structural class specific manner. For each structural class, three best-fitting distributions that describe these distances were obtained from maximum likelihood fit using 'Matlab'.

**Evolution of $3_{10}$ and π-helices**
An investigation of evolution of $3_{10}$ and π-helices was undertaken on twenty largest Pfam families. The broad objective was to examine the changes in the parts of sequence with $3_{10}$ and/or π-structural state over the course long-term evolution. To investigate this, study was carried out in the pairs of proteins with different evolutionary distances, within these Pfam families, taken one at a time. Since such changes should be more clearly revealed in the proteins with different evolutionary distances within a family; Pfam provided an ideal framework for comprehensive study. We obtained sequences for the proteins with solved crystal structures available in PDB for each family. Protein families having less than four structures available or having no RH were excluded from analysis. A flowchart of evolutionary analysis can be found in **Fig. 8.** Detailed descriptions of the families were taken into consideration after removing the anomalies, an account of which has been provided in **Supplementary Table 1**.

Multiple sequence alignment was performed for remaining families. Evolutionary distances between each pair of the proteins within a family were computed from this multiple sequence alignment using MEGA (Molecular Evolutionary Genetics Analysis) [74]. Distance calculation requires at least one common aligned site in the multiple sequence alignment. Hence, only the sequences with comparable lengths were considered. Since substitution rate is known to vary with amino acid pair, we computed JTT (Jones-Taylor-Thornton) distances [75] for evolutionary distance comparison. To ensure the best possible result, substitution rate was assumed to follow gamma distribution with shape parameter 2.4 [76]. Distance comparison studies yielded a symmetric matrix of pair-wise evolutionary distances for each family. For each pair of proteins in a family in the evolutionary distance matrix (an upper-triangular matrix excluding self), the number of substitutions of a $3_{10}$-helix with any other secondary structure in the multiple sequence alignment of the proteins was counted. Suitable substitution criterion was decided after a survey of the evolutionary distances and alignments as described below.

Any one of the sequences from the pair under consideration was chosen as the reference sequence. Locations of the $3_{10}$-helices (assigned by DSSP) were mapped on aligned sequences. If any of the $3_{10}$-helices in the reference sequence occurred within a 3-residue range in the other sequence, it was considered to be on the same sequence coordinates. This 3-residue buffer was used to account for the possible insertion/deletion in another region that might affect the position of the $3_{10}$-helix under consideration. The event of absence of a $3_{10}$-helix within this 3-residue range in the reference sequence can either be due to a replacement of the $3_{10}$-helix or incorporation of a new $3_{10}$-helix in one of the sequences. Final decision to select between these two possibilities was taken on the basis of a study of evolutionary distance, under the assumption that with the increase of evolutionary distance, the tendency to lose a RH in a pair of aligned proteins also increases. From a manual analysis of distances and occurrence patterns, we found that the mean evolutionary distance (for the family under consideration) can be used as a consistent and efficient parameter to choose between replacement and new RH insertion. Therefore, as a logical continuation, when the evolutionary distance of the pair under consideration was found to be greater than the mean distance for that family, it was considered to be a case of replacement. The secondary structure replacing the RH was noted in this case. Whereas, when the evolutionary distance of the pair was found to be less than or equal to the mean distance for family, a case of insertion of a new RH was registered. This procedure was repeated for all the pairs of proteins in all Pfam families, where at least one of the sequences contains at least one $3_{10}$ or π-helix. An example multiple sequence alignment and evolutionary distance can be found in **Supplementary Material** for family PF00033 (Cytochrome b).

**Table-1**

|  | $3_{10}$-helices | | π-helices | | α-helices | |
|---|---|---|---|---|---|---|
|  | Absolute number | Normalized percentage | Absolute number | Normalized percentage | Absolute number | Normalized percentage |
| **Disallowed** | 1575 (**457**) | 1.29 (**1.54**) | 7 (**2**) | 1.23 (**2.00**) | 7906 (**2954**) | 0.75 (**0.75**) |
| **Generous** | 1462 (**5**) | 1.20 (**0.02**) | 18 (**1**) | 3.17 (**1.00**) | 4225 (**153**) | 1.31 (**0.02**) |
| **Allowed** | 11147 (**3026**) | 9.12 (**10.24**) | 143 (**15**) | 25.22 (**15.00**) | 26210 (**10190**) | 2.50 (**1.39**) |
| **Core** | 108059 (**26046**) | 88.40 (**88.19**) | 399 (**82**) | 70.37 (**82.00**) | 1011239 (**717102**) | 96.35 (**98.18**) |

**Table 1.** Occurrences of residues in different regions in Ramachandran plot for $3_{10}$, π and α-helices. Numbers in bold represent numbers of non-glycine, non-proline residues.

**Figures Legends:**

**Figure 1.** The Ramachandran Maps for (A) $3_{10}$-helices and, (B) $\pi$-helices.

**Figure 2.** Observed and predicted mean occurrence of $3_{10}$ helices for all non-redundant proteins. For (A) all-$\alpha$ (B) $\alpha/\beta$ (C) $\alpha+\beta$ (D) membrane (E) multi-domain and (F) small proteins. "Real" trends are assigned by DSSP, "predicted" trends are obtained from present probabilistic model. The range of ordinate is 0 to 1 (total range of probability); however, only the parts with significant differences between predicted and observed magnitudes are shown here. Abscissa describes number of cases (see text for description).

**Figure 3.** Observed and predicted mean occurrence of $\pi$-helices for all structural classes.

**Figure 4 -** Modeling the inter-arrival interval for $3_{10}$-helices on protein sequences for (A) all-$\alpha$ (B) $\alpha/\beta$ (C) $\alpha+\beta$ (D) membrane (E) multi-domain and (F) small proteins. Histogram represents the probability density of sequence intervals between consecutive $3_{10}$-helices. Different colored lines represent fitted distributions.

**Figure 5.** Modeling the inter-arrival interval for $\pi$-helices on protein sequences for all structural classes.

**Figure 6.** (A) Percentage of cases of replacements retention and 'new' insertions of $3_{10}$-helices in different Pfam families. (B) Percentages of different secondary structures ($\alpha$-helices, $\beta$-sheets and loops) replacing $3_{10}$-helices in different Pfam classes.

**Figure 7.** Cases of replacements, retention and 'new' insertions of $3_{10}$-helices with increasing evolutionary distance for all studied Pfam families. Evolutionary distances have been divided into bins of 0-1, 1-2 and so on. The last bin contains cases with distances greater than 10.

**Figure 8.** Flow-chart of methodology to identify evolutionary fate of RHs.

**Table 1.** Occurrences of residues in different regions in Ramachandran plot for $3_{10}$, $\pi$ and $\alpha$-helices. Numbers in bold represent numbers of non-glycine, non-proline residues.

**Supplementary Material** (A) Multiple sequence alignment for PDB structures from family PF00033 (Cytochrome b). $3_{10}$-helices have been highlighted in red color and three cases of replacement, retention and new $3_{10}$-helix insertion are marked by green, yellow and red squares respectively. (B) Evolutionary distance matrix for Cytochrome-b family used to assign different evolutionary scenarios mentioned above. Average distance between family members is 0.939.

**Supplementary Table 1.** Family ids and details of 20 largest PFam families considered for evolutionary analysis.

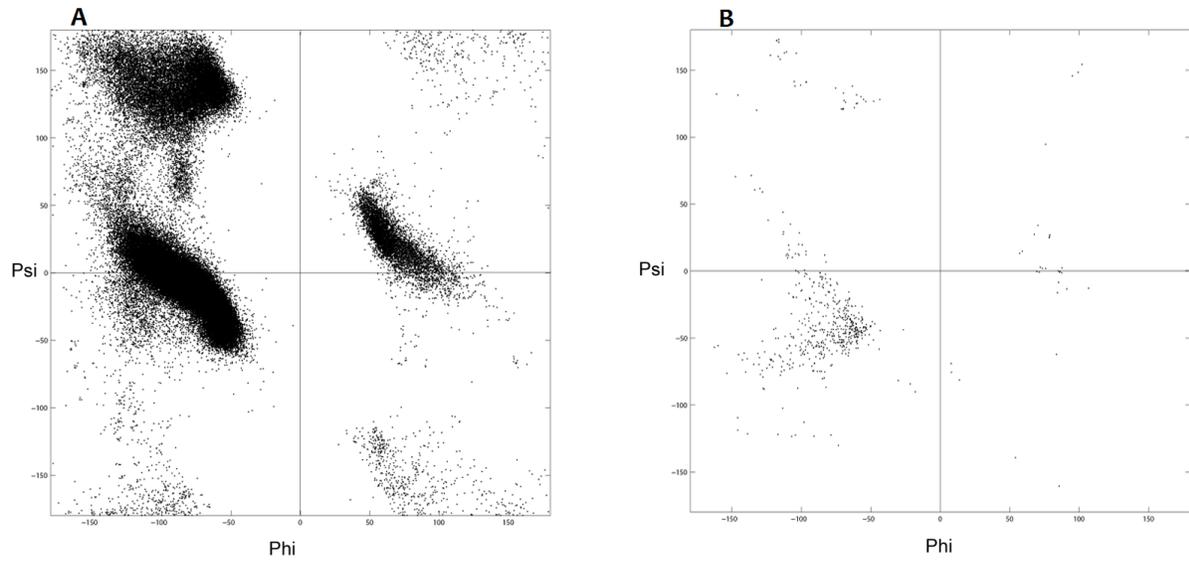

**Figure-1**

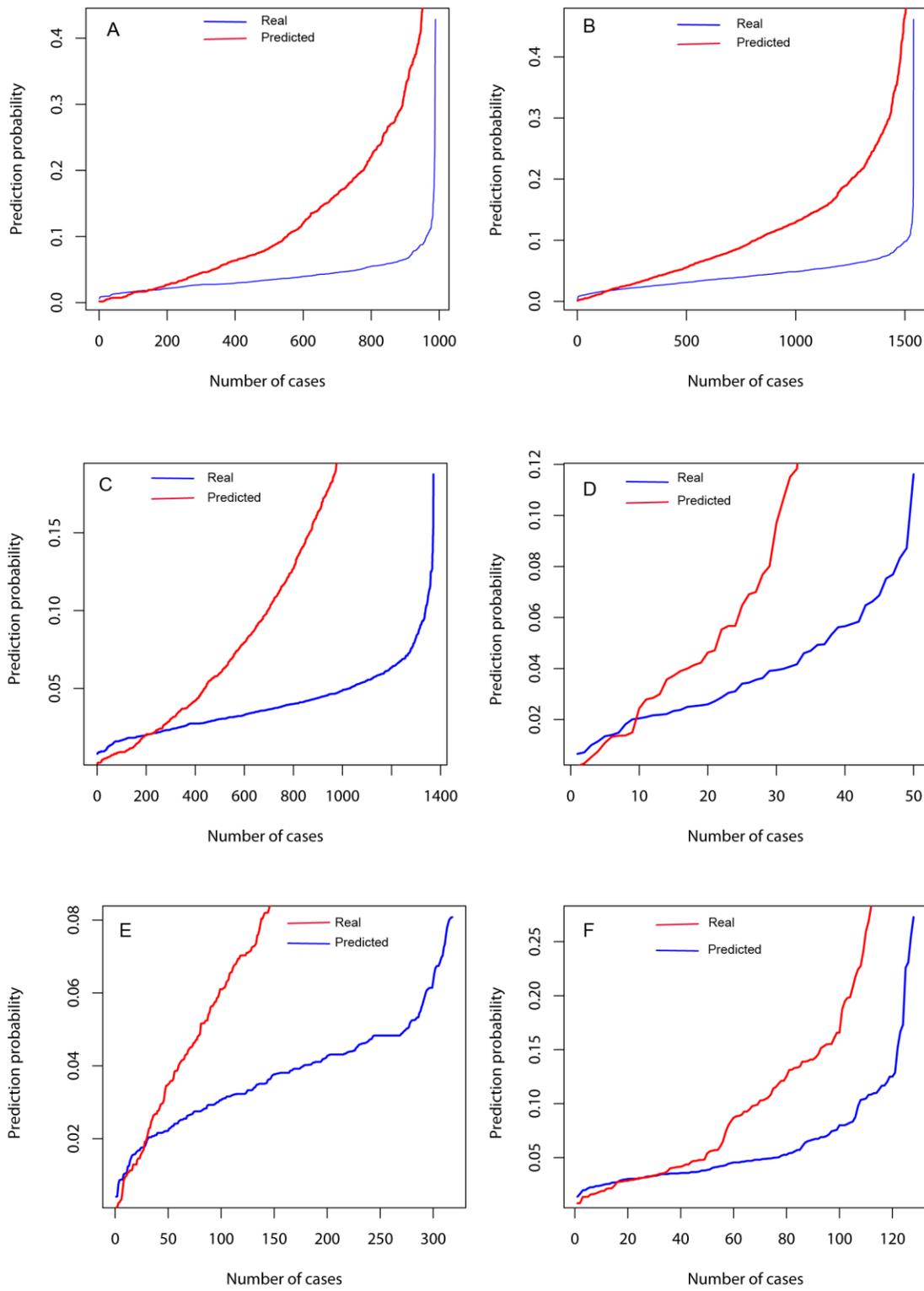

**Figure-2**

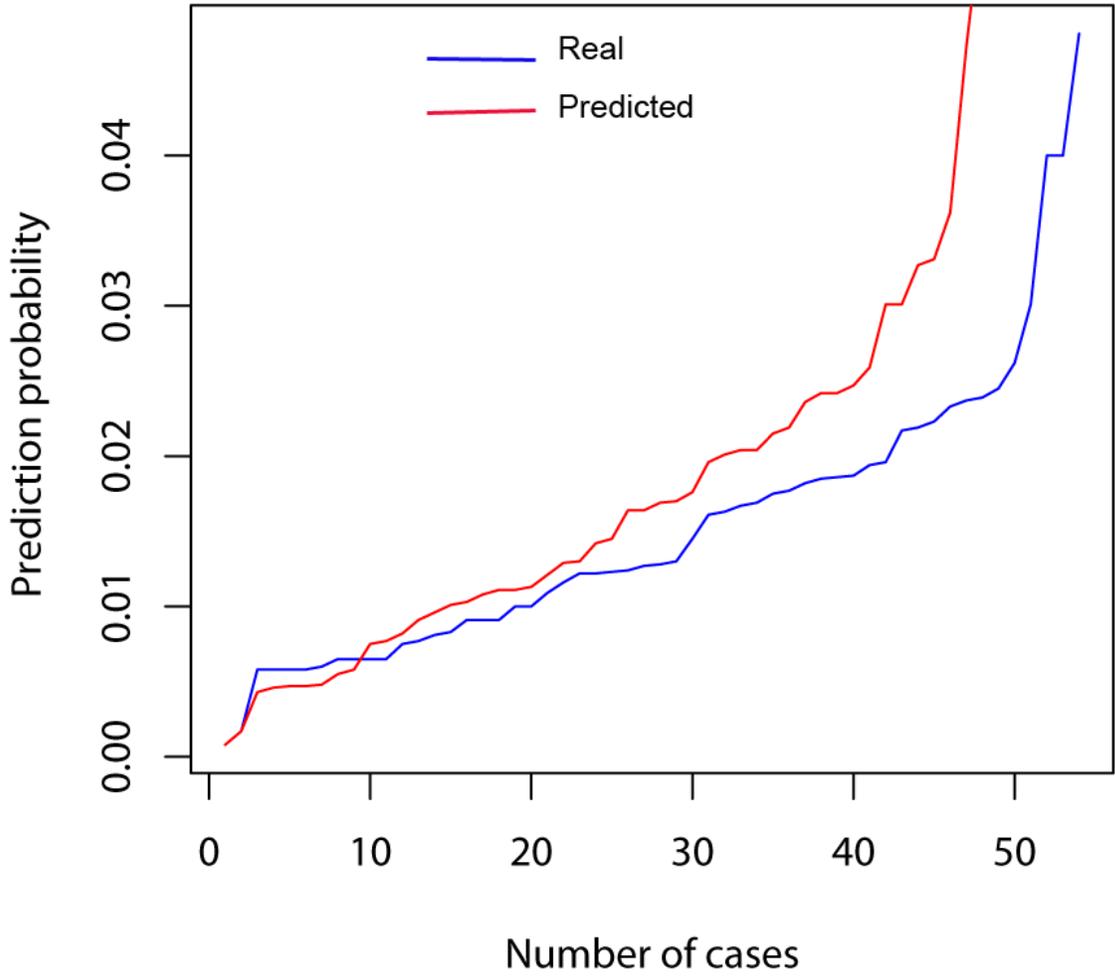

Figure-3

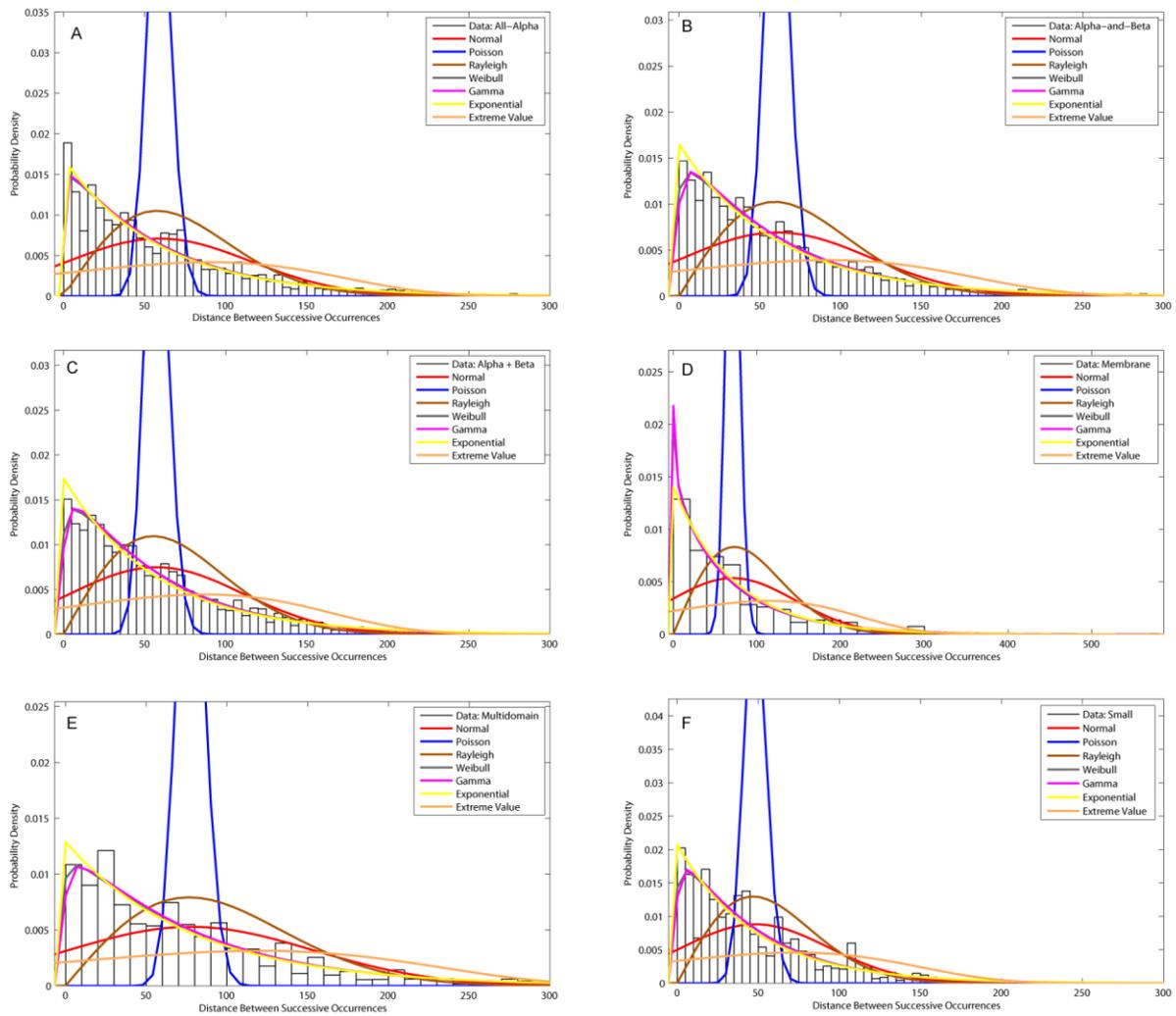

**Figure-4**

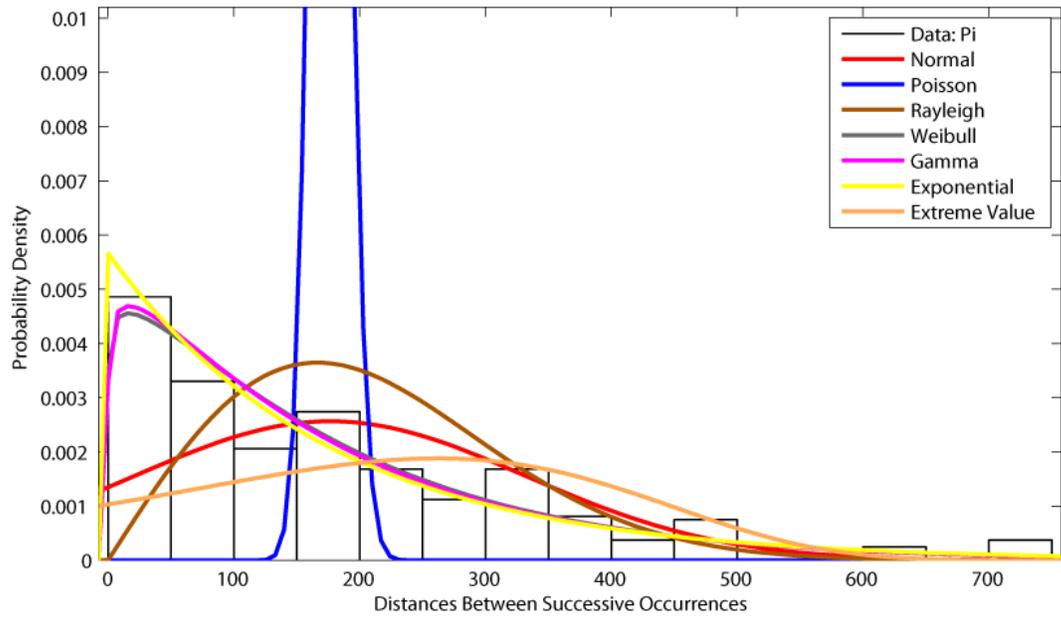

**Figure-5**

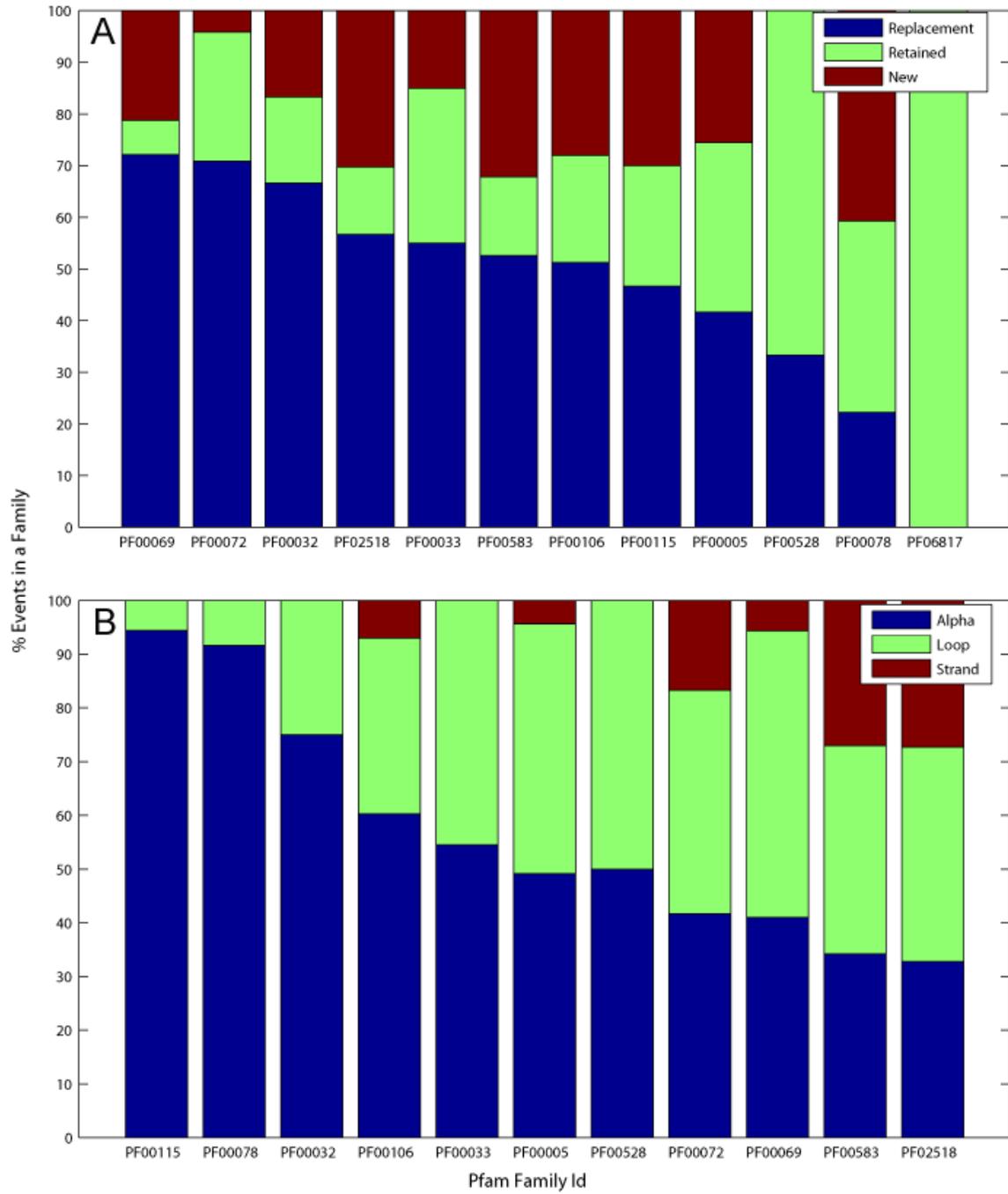

**Figure-6**

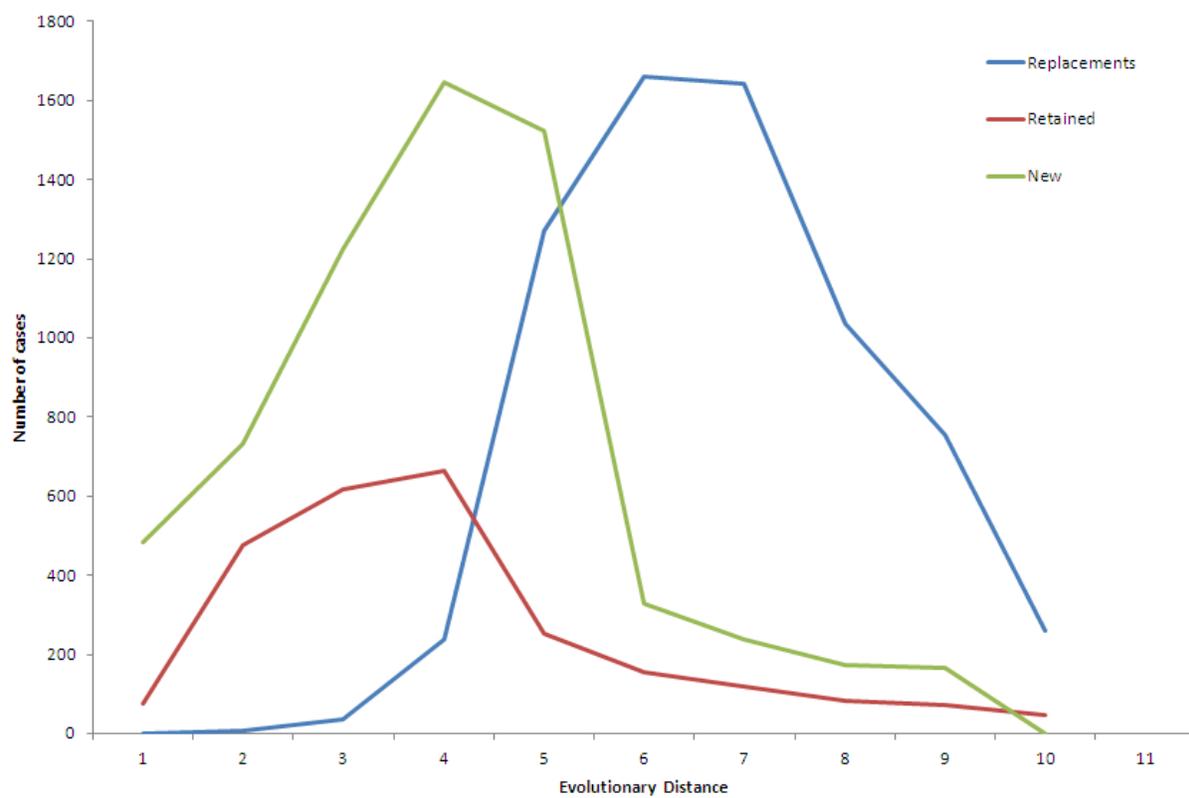

**Figure-7**

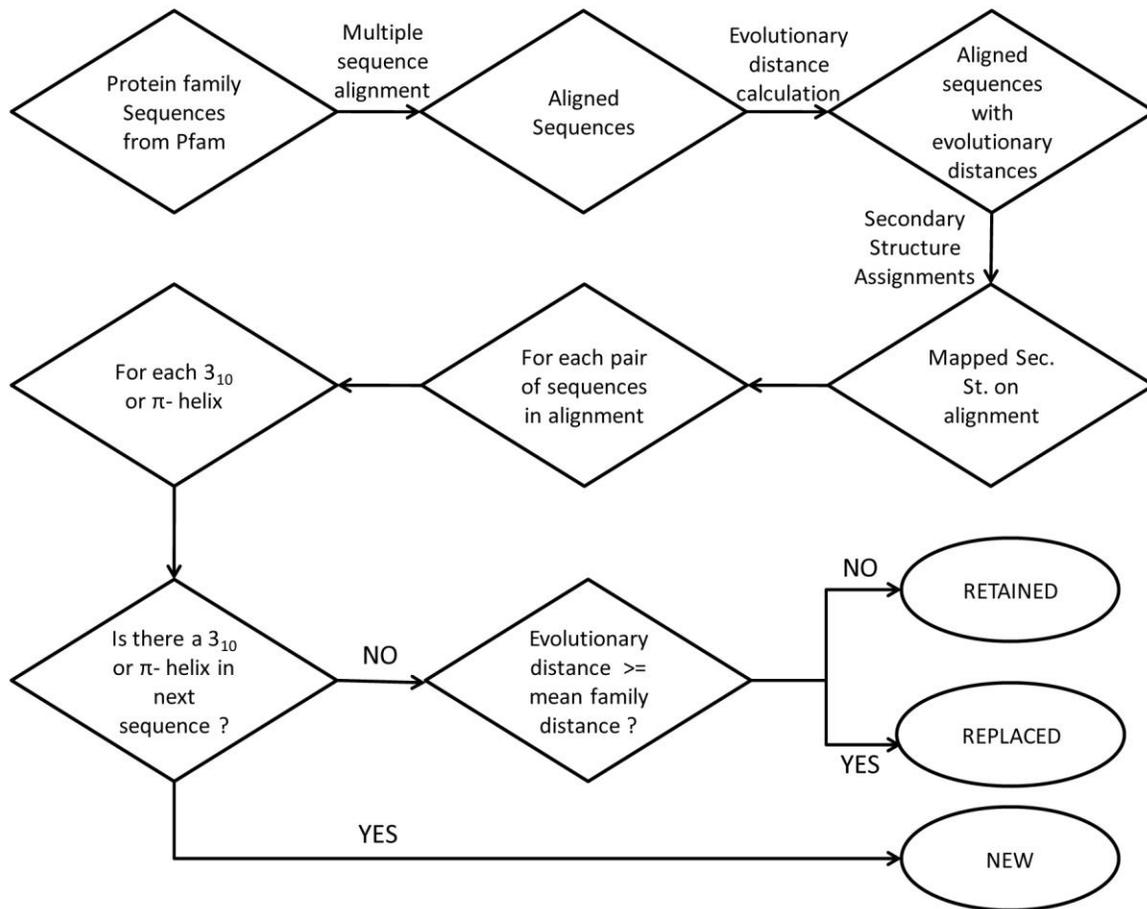

**Figure-8**